# Plasmonic Airy Beam on Metal Surface

**L. Li, T. Li★, S. M. Wang, and S.N. Zhu★**

National Laboratory of Solid State Microstructures,

College of Physics, College of Engineering and Applied Sciences, Nanjing University,

Nanjing 210093, China,

**Optical Airy beam, as a novel non-diffracting and self-accelerating wave packet, has generated great enthusiasm since its first realization in 2007, owing to its unique physics and exciting applications. Here, we report a new form of this intriguing wave packet - plasmonic Airy beam, which is experimentally realized on a silver surface for the first time. By particular diffraction processes in a carefully designed nano-array structure, a novel planar Airy beam of surface plasmon polariton (SPP) is directly generated and a structural dependent phase tuning method is proposed to modulate the beam properties. This SPP Airy beam is regarded as a two-dimensional (2D) subwavelength counterpart of the 3D optical Airy beam in free space, allowing for on-chip photonic manipulations. Moreover, it possibly suggests a breakthrough in recognition of this unique wave packet in a polariton regime after its previous evolution from free particle to pure optical wave.**

★Corresponding authors, E-mails: taoli@nju.edu.cn, zhusn@nju.edu.cn

Website: http://dsl.nju.edu.cn/dslweb/en-index.html

Airy wave packet is the only nontrivial 1D solution for a wave propagation maintaining the non-spreading property, which was deduced from Schrödinger equation in quantum mechanics for a free particle[1]. Since its recent observation in optics[2], intensive studies have been carried out on its novel properties, such as self-accelerating[3], ballistic dynamics[4], self-healing[5,6], as well as the recent nonlinear generation[7] and possible applications[8-10] (e.g., particle clearing, curved plasma channel). Up to date, these experimentally achieved optical Airy beams were all generated in 3D free space. An intuitive extension of this unique wave packet to two dimensions would possibly forecast more fascinating physics and applications, especially as it is accommodated in the subwavelength plasmonics, which is another intriguing field nowadays[11-15]. As has been envisioned in a recent theoretical study[16], plasmonic Airy beam would provide effective means to route energy over a metal surface between plasmonic devices. Moreover, since surface plasmon polariton is an elementary excitation that can be regarded as a kind of quasi particle merging the optical field and electron oscillation, the realization of SPP Airy beam would possibly indicate further extension of Airy wave packet from the pure light to the quasi particle system.

According the nature of Airy wave packet[1], a 3/2 power phase modulation along the lateral dimension of beam is required[17,18]. It is commonly modulated to 3 power phase type by a mask on an incident Gaussian beam with a followed Fourier transformation in generations of free space Airy beams[2-10]. This 3/2 phase requirement is inherited for an SPP Airy beam, although the field form is modified[16]. However, the conventional method tends to be hardly adopted in a surface regime due to the complex transformation process and large spatial expense. Although alternative approach was conceived by coupling a free space generated Airy beam into a planar plasmonic one[16], it will inevitably bring other severe obstacles (e.g., coupling process, the mismatched characteristics in different scales) and remains great challenges.

Here, we report the first realization of plasmonic Airy beams on a silver surface at visible wavelength, which is accomplished by particular diffraction processes with respect to a carefully designed nanocave array on the metal film. This experimentally achieved plasmonic Airy beam intuitively reveals the non-spreading and self-bending property over a considerable long distance (~50 μm), demonstrating the capacity of a transversely self-confined SPP beam in a planar dimension with lower propagation loss. Furthermore, the proposed diffraction approach by designable nanostructures has exhibited it flexibility in beam tailoring, which may significantly broaden the study in manipulation of the SPP waves in planar dimension.

**Results**

**Direct generation of SPP Airy beam by graded nanocave array.** Using periodic array on metal surfaces to manipulate SPP propagations has achieved great success in recent years[19-22]. Actually, these approaches used to change the SPP propagations can also be interpreted as a phase modulation, which is in coincidence with the descriptions in diffraction optics. In principle, it is quite possible to use inhomogeneous array to change a linear phase of incident SPP into a nonlinear one. In this regard, the Airy beam required 3/2 phase modulation is highly expected by diffractions in non-periodic array system. The scheme of our design is shown in Fig. 1. On the surface of a silver film (with $SiO_2$ as the substrate), a groove grating is used to couple a He-Ne laser beam ($\lambda$=632.8nm) into an in-plane propagating SPP wave, which subsequently incidents into a non-periodically arranged nanocave array. By appropriate arrangement, diffracted SPP waves from nanocaves will interfere and ultimately build two SPP Airy beams on both sides.

In experiments, the metallic nanostructure was fabricated by the focused ion beam (Strata FIB 201, FEI company), and SPP wave propagation analysis was performed by a home built leakage radiation microscope (LRM) system[23,24] (see methods for details). The inset image in down-right of Fig. 1 is a typical experimental result, which intuitively demonstrates the generation of SPP Airy beams that very analog to the schematic illustration, manifesting the self-bending, non-spreading and multiple lobes. Here, the sample of nanocave array is designed graded in *x*-direction and periodic in *z*-direction. Figure 2a depicts the top view of the graded nanocave array together with a grating coupler, and the recorded SPP beam in the right branch is specifically shown in Fig. 2b for detailed analysis. Subsequently, we performed a theoretical calculation based on the Huygens-Fresnel principle[18], that all nanocaves in the array are considered as sub-sources that radiate cylindrical surface waves (see methods). The calculated beam trajectories (shown in Fig. 2c) are in good agreement with the experiment ones, although they are both imperfect due to limited diffraction elements and non-ideal modulations. To make a quantitative evaluation, beam profiles at different propagation distances (marked in Fig. 2b) are plotted in Fig. 2d, from which a set of Airy-like wave profile are clearly manifested. The main lobe keeps non-spreading property within at least 30 μm distance, which is considerable long for an SPP wave at wavelength of 632.8 nm (λ in free space). A narrow beam width of the main lobe (FWHM ~1 μm) is notably preserved over a long distance. Thanks to this particular instructive interference of diffraction waves, the attenuation of the main lobe of this SPP Airy beam appears much less than a common SPP wave. In this regard, it behaves like a self-confined in-plane waveguide with lower loss and suggests possible applications in guiding SPP waves.

**Phase modulation of SPP Airy beam.** To explain how the SPP Airy beam comes into being, a novel nonlinear phase modulation by diffractions from non-periodic array is introduced. As it is well known, an incident SPP wave will be diffracted (or reflected) into a well defined direction by a designed array determined by the Bragg condition, which can be clearly schemed out in the reciprocal space[25] (see Fig. 3a). However, if this condition is not perfectly satisfied, a preference diffraction will still occur with some sacrifice in intensity (as long as the deviation is not too large), owing to the elongated reciprocal lattice of finite-scale array (see Fig. 3b) that is similar to the X-ray diffraction cases[26]. It is also proved by our experiments (see supplementary information). Therefore, different lattice parameter is able to determine the preference diffraction direction, which can be regarded equivalently to yield an extra phase change of $2\pi$ from every local lattice (in *x* direction here). When a graded array is employed, incident beam will diffract to different directions at different positions according to the local lattice parameters (for this small gradient case $\Delta$=10nm). Thus, we can obtain the corresponding phase evolution from every lattice point in the incident SPP propagation ($x$ direction) as $\phi(x)$= $\phi_0+k_{spp}x-2m\pi$. This phase evolution in turn manifests the gradually changed diffraction directions by graded lattice.

From Fig. 2b, a beaming angle ($\theta\sim20°$) with respect to $z$-axis is found for the main lobe. This means the lattice boundary (line $z$=0) is not the start line of this SPP Airy beam. We can deduce the phase information at a virtue starting line in $\xi$-axis that perpendicular to the tangent of beaming trajectory of the main lobe according to the principle of geometric optics as (see the scheme in Fig. 3c)

$$\phi(\xi) = \phi(x) - k_{spp}b = 2m\pi + k_{spp}x - k_{spp}b, \tag{1}$$

where

$$b = -\frac{x\tan(\theta_0)}{\cos(\theta_x)+\tan(\theta_0)\sin(\theta_x)}, \quad \xi = \frac{x}{\cos(\theta_0)(1+\tan(\theta_x)\tan(\theta_0))}, \tag{2}$$

and $\sin(\theta_0) = \frac{\lambda_{spp}-a_0}{a_0}, \sin(\theta_x) = \frac{\lambda_{spp}-a_x}{a_x}$, $a_x$ is defined as the local lattice determined by the mean value of two distances before and after the lattice of $x$. According to the experimental result of the position of original point of $O$ ($a_0\sim450$ nm) and initial angle ($\theta\sim20°$), we calculated the transformed phase $\phi(\xi)$ shown in Fig. 3d together with the results of 1.4, 1.5 and 1.6-power phase modulations. It is clearly seen that the deduced data from the graded array matches the 3/2 power relation considerably well. It well explains the outcome of Airy-like beam. In addition, the intensity of diffraction from every local lattice is dominated by the matching condition, i.e., the better Bragg condition satisfied, the stronger diffraction formed. Thus, the location of the main lobe is expectedly near the match point, which is in coincidence with the Airy function.

**Designable generation of SPP Airy beam.** Based on above phase modulation method, SPP Airy beam with a defined beaming direction (for the main lobe) can be generated by a proper non-periodic array. With a defined beaming angle of $\theta$, the corresponding phase along the $x$ axis can be retrieved as

$$\psi(x) = -\frac{2}{3}\left(-\frac{\xi}{\xi_0}\right)^{3/2} - \frac{\pi}{4} - k\frac{\xi \sin\theta}{\cos(\theta - \theta_\xi)}, \tag{3}$$

where $x = \xi \cos(\theta) + \xi \sin(\theta)\tan(\theta - \theta_\xi)$, $\theta_\xi = \arcsin(\partial_\xi \phi(\xi))$, $\phi(\xi)$ is the phase satisfying the Airy function, and $\xi_0$ is a constant determines the acceleration of Airy beam. According to the equivalent phase by diffraction $\phi_m(x) = kx + 2m\pi$, we can deduce the location of the $m$th diffraction unit by solving $\phi_m(x) = \psi(x)$, and ultimately retrieve the arrangement of nanocave array. Fig. 4a-4d are the designed array data and calculated results of the SPP Airy beam with the angle of $\theta=0$ and -7°, respectively. The corresponding experimental results are subsequently shown in Fig. 4e and 4f, which agree well with the calculated ones. By carefully examining these beaming profiles, a set of upper diffraction branches with considerable strong intensities are exhibited besides main Airy beams. It is actually due to another matched condition corresponding to the reciprocal $G_{-2,1}$. Here, of importance is that SPP Airy beams are realized in a designable way by proper phase modulation, in which the non-spreading, self-bending properties are well demonstrated.

**Discussion**

According to above demonstration of SPP Airy beams achieved by graded arrays (from the fixed gradient case to the designed ones), an artificial modulation of the diffraction phase is well proved to have the capacity to manipulate the SPP wave propagations. However, another characteristic of the Airy beam - lateral intensity modulation - is not addressed, because it is rather complicated for this non-periodic case and remains a problem to be further explored. Fortunately, the graded system is commonly able to build a localized wave packet at its propagation end (around the matched condition)[27], which usually has asymmetric profile that considerably analog to the intensity envelope required by the Airy function. As for more precise intensity modulation (e.g., the -1/4 power relation[1]), we believe it would be accomplished or improved by carefully tuning the diffraction elements (e.g., variable diameter, depth or shapes of the nanocaves, or nano bulbs), as well as modifications of the array periods in the other dimensions. Even though, as has been mentioned that the phase modulation is the critical factor[17,18], we have successfully achieved the SPP Airy beam on silver surface in a controllable way.

For further perspective, the parameters of the nanocave arrays are designable with respect to the defined wavelength of SPPs. It is reasonable to suggest a frequency dispersive array structure to generate multi-frequency (or colorful) SPP Airy beams by a single design, which may indicate good functionality in planar plasmonic manipulation and integrations. Moreover, the proposed method by diffraction process with nanocave array exhibits more flexible than traditional methods, which may be used to generate other beam forms of SPP as well as some other waves as expected. Since the SPP is a kind of polariton with a mixed character of particle and wave, this type of Airy would probably indicate the existence of some other quasi-particles (e.g., phonon, magnon, etc). In this sense, our work would open a new avenue for further exploration of such a unique wave packet. In addition, the SPP Airy beam detected by LRM system is a direct observation of the unique Airy beam trajectory that cannot be obtained in previous optical beams in free space.

In conclusion, we have developed a new method to design and experimentally demonstrate, for the first time, the SPP Airy beam on a silver surface. The revealed SPP Airy beam exhibits non-spreading property with about 1 μm lateral confinement for the main lobe over a long distance at visible wavelength, which has implications in SPP manipulation and other related field (e.g. arranging nanoparticles in nanoscale). Furthermore, the method based on the diffraction effect allows for controllable modulations on the established plasmonic Airy beam almost at will, which may have more general applications in the wave-front tailoring as well as developing new kind of photonic or plasmonic structures and devices.

**Methods**

**Fabrication and LRM optical analysis.** The nanocave array sample was fabricated by focused ion beam (FEI Strata FIB 201, 30keV, 11pA) milling on a 60nm thickness silver film, which has been deposited on a 0.2mm-thinkness SiO$_2$ substrate. The propagation of SPP waves was analyzed by the approach called leakage radiation microscope (LRM) system[23]. Briefly, it works on the radiation mode that leaks from the decaying SPP field through the metal layer into the high refractive substrate (here, $n_{SiO2}$>$n_{air}$), as the metal layer is thin enough. In our experiments, SPP waves were excited by He-Ne laser (632.8nm) focused by a microscope objective (50×, numerical aperture of 0.55) onto the grating coupler. The leakage radiation emitted into the SiO$_2$ substrate from Ag/SiO$_2$ interface was collected by an oil immersion microscope objective (160×, numerical aperture of 1.40). The real space SPP propagation and its Fourier transformation image (i.e., $k$-space) were recorded using a charge-coupled-device (CCD) in the object plane and back focal plane respectively, depending on the location of an auxiliary lens[24].

**Theoretical calculation.** The theoretical calculation is based on the Huygens-Fresnel principle, that all nanocaves in the array are considered as sub-sources that radiate cylindrical SPP surface waves. According to the field form of SPP sub-sources, we can calculate interfered SPP field intensity by summing over the field of all diffracted SPP waves as (due to the TM nature of SPP, we only consider the $y$ component)

$$E_{y,tot}(x,z) = E_0 \sum_{m,n} \frac{1}{\sqrt{r_{m,n}}} \exp[-ik_{spp}(d_{m,n} + r_{m,n}) - \alpha(d_{m,n} + r_{m,n})]\exp(i\omega t),$$

where $k_{spp}$ is wave vector of SPP, $d_{m,n}$ is the distance from the initial position of SPP incidence to the ($m$, $n$) lattice and $r_{m,n}$ is the distance from the lattice point to a location of ($x$, $z$). The decay coefficient $\alpha$ is fitting parameters corresponding to the attenuation of SPP wave propagating on the planar surface. Therefore, the intensity of interfered SPP wave beaming from the nanocave array can be obtained from $\left|E_{y,tot}(x,z)\right|^2$.

**Acknowledgements**

The authors thank Dr. C. Zhang and Dr. P. Xu for beneficial discussions in theoretical analyses. This work is supported by the State Key Program for Basic Research of China (Nos. 2009CB930501, 2010CB630703 and 2011CBA00200) and the National Natural Science Foundation of China (Nos. 10974090, 60990320 and 11021403),

**Author Contributions**

L. Li, T. Li, S. M. Wang, and S. N. Zhu

T.L. and S.N.Z. supervised the study. T.L. and L.L. conceived and designed the experiments. L.L. fabricated the sample and performed the optical analyses. L.L. and S.M.W. performed the numerical simulations. T.L. and L.L. analyzed the results and wrote the paper with assistance from S.M.W. and S.N.Z.

**Additional Information**

The authors declare no competing financial interests.

Supplementary information accompanies this paper is provided online.

Correspondence and requests for materials should be addressed to T.L.

**Figure Legends**

**Figure 1 | Schematic of direct generation of the SPP Airy beam.** A laser beam is coupled into in-plane propagating SPP wave by grating and incidents into a non-periodically arranged nanocave array. Two SPP Airy beams are formed on both sides of the array by diffraction processes. Inset is a typical experimental result of SPP Airy beam examined by the leakage radiation microscope system.

**Figure 2 | SPP Airy beam generated in experiment and theoretical calculation. a,** Top view of the graded nanocave array sample fabricated by the focused ion beam, where the lattice parameter is graded in *x*-dimension ($a_x$ from 420nm to 780nm, grads $\Delta$=10nm) and period in *z*-dimension is $p_z$=620nm. **b,** Experimentally achieved SPP beam trajectories and **c,** Calculated one. **d,** Beam profiles of the experimental SPP beam in different propagation distances marked with dash line in panel (**b**).

**Figure 3 | SPP diffraction schemes and phase evolution of nanocave array. a, b,** Ewald construction for SPP diffraction direction with (**a**) Bragg condtion is satisfied and (**b**) Bragg condition is not perfectly satisfied with limit diffraction elements in *z* axis (elongated reciprocal lattices in *z* axis are indicated). $k_{spp,i}$ and $k_{spp,d}$ are the incident and diffracted SPP wave vectors, respectively. $G_{0,1}$ and $G_{1,0}$ are two basic vectors of the reciprocal lattice. According to the schemes, the diffraction directions are determined by the latice parameter in *x* axis for both (**a**) and (**b**). **c,** Scheme of the phase transformation from the *x* axis to a virtue $\xi$ axis, which can be designed with respect to the beaming angle $\theta$ for the main lobe of SPP Airy beam. **d,** Deduced phase distrubutions in the

starting $\xi$ axis together with the 1.4, 1.5, 1.6-power phase modulations.

**Figure 4 | Designable generation of SPP Airy beam. a, b,** Designed non-periodic lattice in *x*-dimension (scaled with respect to the bottom of the array) for (**a**) horizontal beaming $\theta=0°$ with $p_z$=640 nm, $\xi_0$=1.08 and (**b**) down-inclined beaming $\theta$=-7° with $p_z$=650 nm, $\xi_0$=1.33. **c, d,** Caculated results and **e, f,** Experimental results respectively.

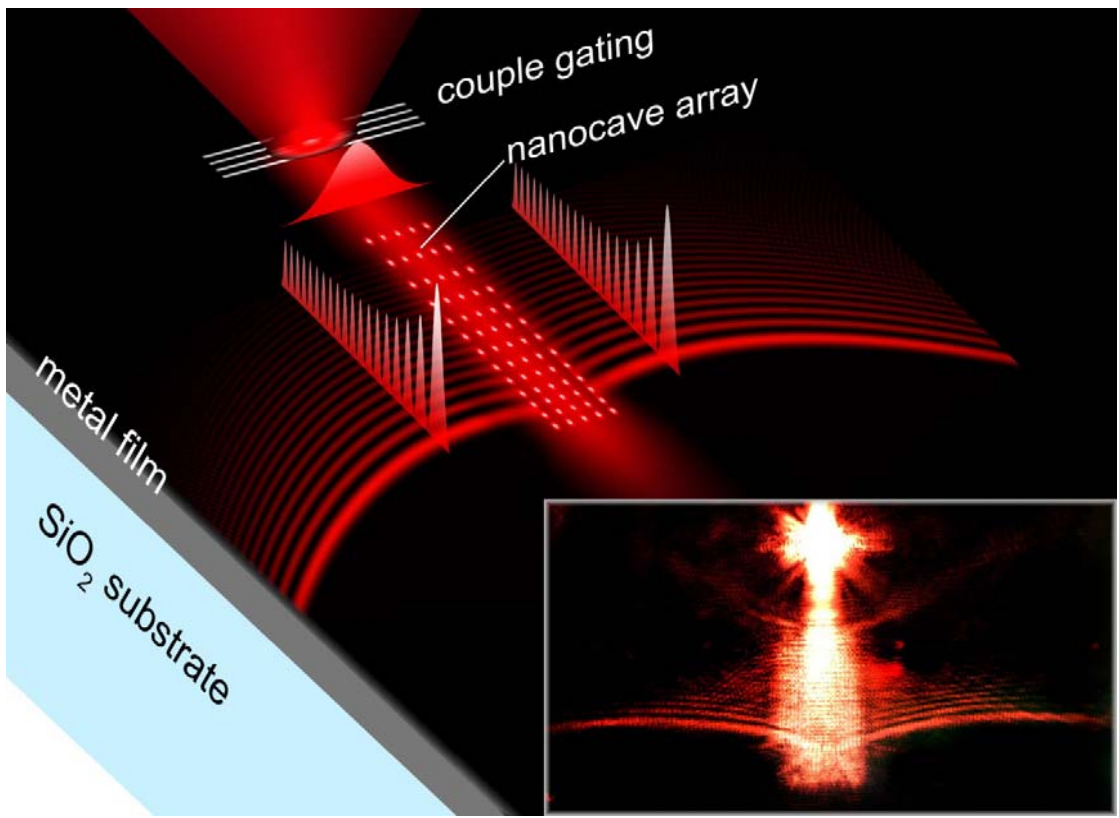

Figure 1

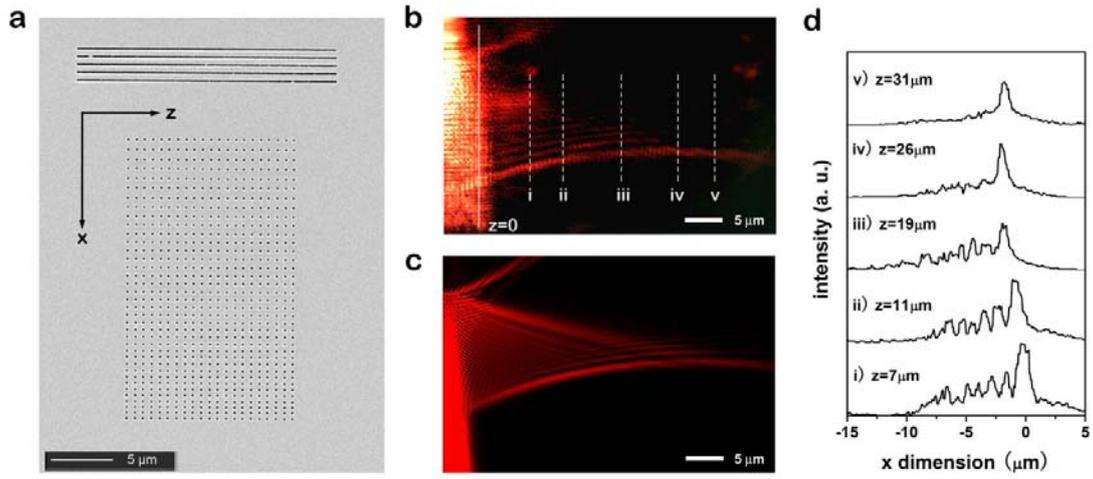

Figure 2

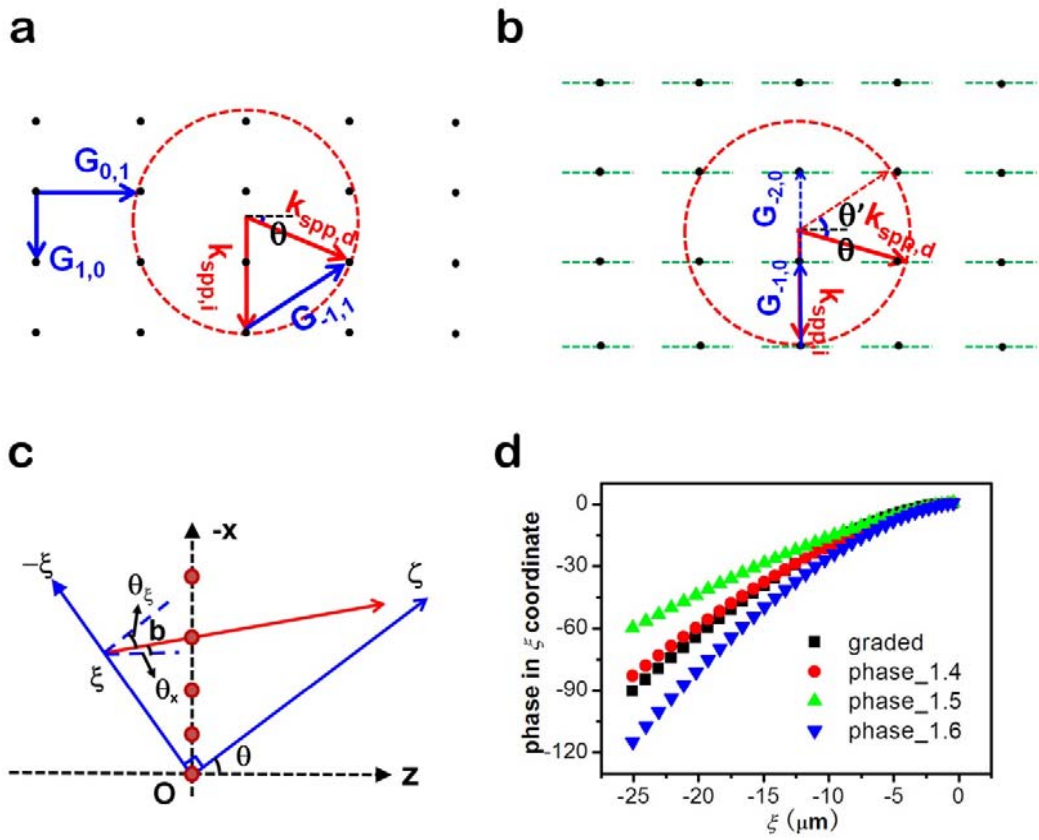

Figure 3

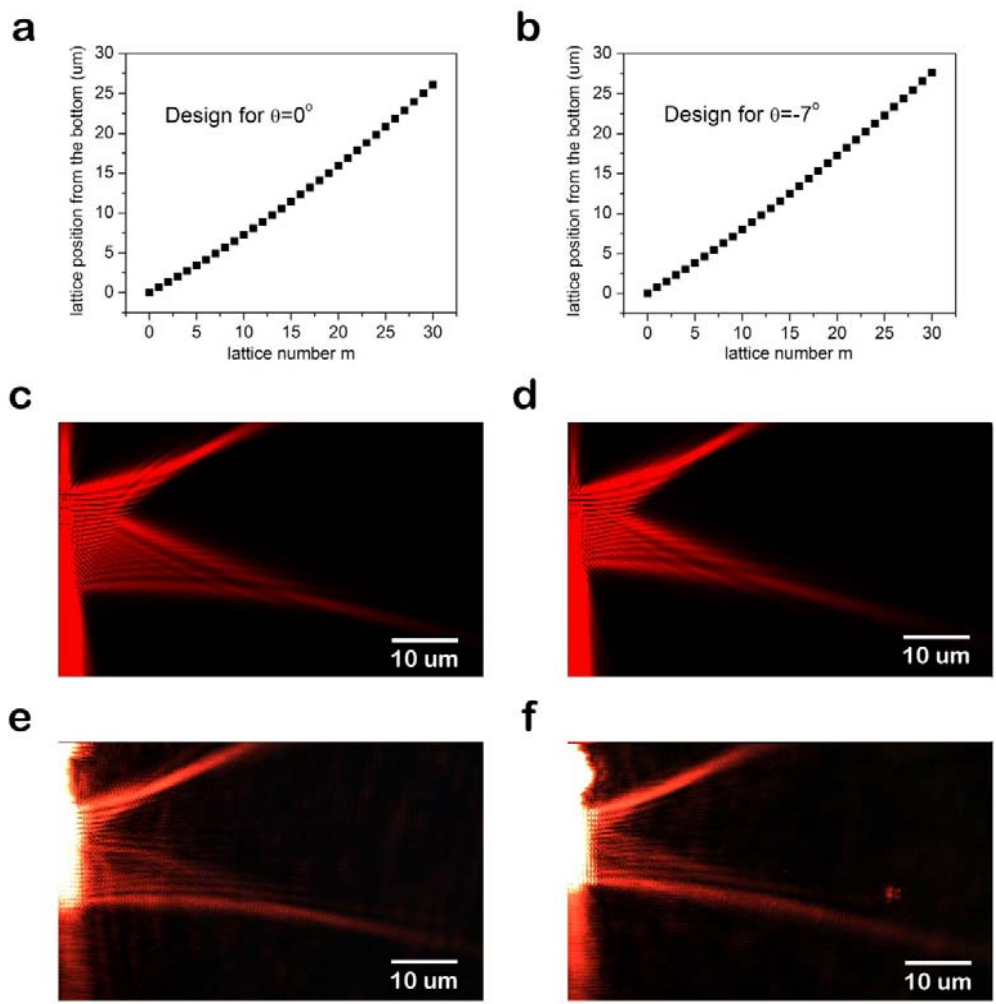

Figure 4

# Supplementary information

**Preference diffraction by nanocave array**

The interaction between the SPP wave and the 2D array (schemed in Fig. S1a) is conveniently described in reciprocal space (Fig. S1b). When Bragg condition is satisfied, an incident SPP wave will be reflected to a preference direction, which is visualized in the Ewald's circle construction. $k_{spp,i}$ and $k_{spp,d}$ are the incident and diffracted SPP wave vectors, respectively. $G_{0,1}$ and $G_{1,0}$ are two basic vectors of the reciprocal lattice. The first order match condition is to be considered in our schematic, so that the Bragg condition is

$$\vec{k}_{spp,d} - \vec{k}_{spp,i} = \vec{G}_{-1,1} \tag{1}$$

as shown in Fig. 1b. Therefore, the allowed diffracted direction can be defined by the angle of

$$\theta = \arcsin\left(\frac{G_{1,0} - k_{spp}}{k_{spp}}\right). \tag{2}$$

However, this Bragg condition cannot be satisfied all the time for an arbitrary designed array, i.e., no other reciprocal lattice rightly locates on the Eward's circle. Notwithstanding this mismatch will destroy the uniform diffraction beam to some extent, the allowed diffraction beam will still exist if the deviation is not too large. Since the incident SPP beam is only about 2~3 μm in width in our experiments covering 3~5 scattering periods in transverse dimension ($z$ direction for the incident SPP beam), the reciprocal lattice will be elongated in *z* direction correspondingly analog to the cases in X-ray diffraction by rod or sheet samples [see Ref. 26]. Diffraction conditions are thus satisfied where the rods intersects the Ewald's circle. Therefore the case of diffraction from Bragg condition (1) approximately reduces to the 1D form:

$$k_{SP,dx} - k_{SP,ix} = G_{1,0}. \qquad (3)$$

The allowed diffraction is still satisfied with the angle of $\theta = \arcsin\left(\frac{G_{1,0} - k_{spp}}{k_{spp}}\right)$ and may sacrifices in intensity compared with the perfectly satisfied Bragg condition. So in both Fig. S1b and S1c, $P_x - a = \lambda_{spp}$ is always satisfied that can be deduced in the real space as well (see Fig. S1a). Then, the diffracted SPP beam can be regarded starting from every lattice point in *x*-direction each has an extra $2\pi$ difference in phase to its neighbors. In this regard, the relationship between the preference diffracted beam (beaming angle) and the lattice property of nanocave array (i.e., locations of every lattice point) is established in description of a spatially defined phase modulation. In other words, we can artificially design the beaming angle of diffracted SPP beam as well as the local phase by tuning the lattice parameter, thanks to the validation of Eq. (2) in the slightly mismatched Bragg condition.

This diffraction property was subsequently confirmed by our experiments. A series of samples of nanocave arrays with various periods in *x* direction ($P_x$ is from 360 to 1100 nm, $P_z$ is fixed at 700 nm, 21 samples in total) were fabricated. In order to make a precise measurement of the diffraction beams, two identical arrays were designed in mirror symmetry with respect to a groove grating in *z* axis, which is used to coupling the 632.8 nm laser beam into an in-plane SPP wave, see Fig. S2a. The diffraction properties of all samples were analyzed by the LRM system systematically. Here, we provide LRM results of two typical samples in Fig. S2, where the real space images of SPP diffractions of samples with $P_x$= 390nm (S2a) and $P_x$= 790nm (S2c) and the corresponding Fourier images (Fig. S2b and S2d) are clearly exhibited. The *k*-space images in the Fourier planes appear in good agreement with cases schemed in Fig. S1c and S1d, confirming our theoretical analysis. Moreover, for the sample of $P_x$= 790nm, another matched condition is revealed for a higher ordered reciprocal vector $G_{-2,1}$ (shown in Fig. S1d and S2d), it consequently results in a strong diffraction beam in this order, as shown in Fig. S2c. It also confirms our prediction of the intensity sacrifice from the mismatch. By carefully measure the diffraction angles of all samples in their Fourier plane, we obtained the whole experimental data, which are in extremely good agreements with the calculated ones from Eq. (2) as shown in Fig. S3.

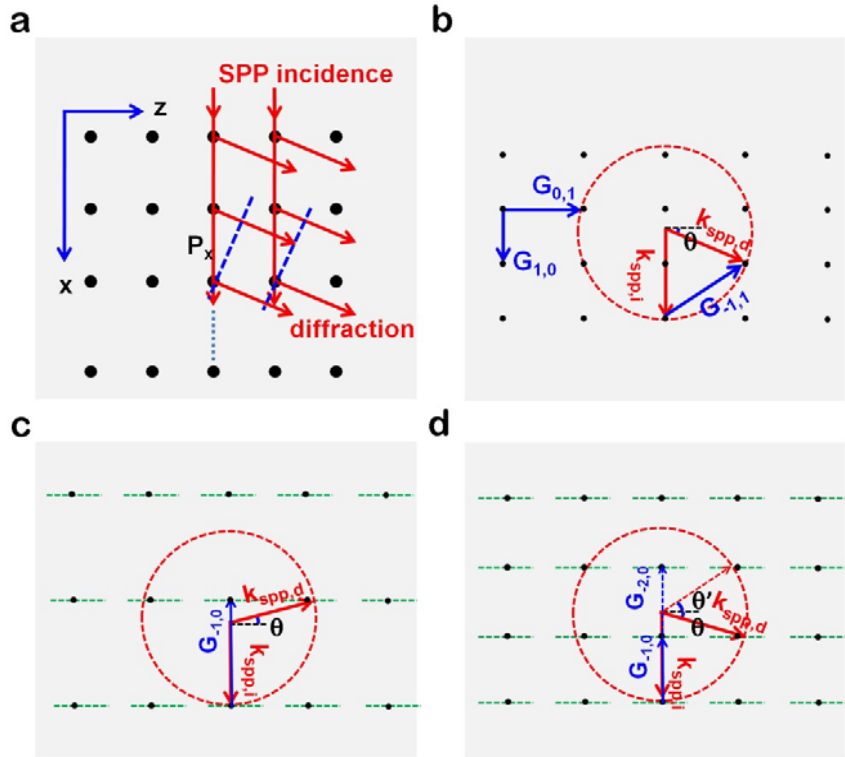

**Figure S1:**

(a) Schematic of incident and diffracted SPP wave in a 2D lattice; (b) The corresponding reciprocal space with a perfectly satisfied Bragg condition, where reciprocal $G_{-1,1}$ rightly locates in the SPP Eward's circle; Reciprocal patterns of slightly mismatched conditions for a decreased (c) and increased (d) of $P_x$, where a higher ordered diffracted $k$ vector is indicated in (d) with the angle of $\theta$'.

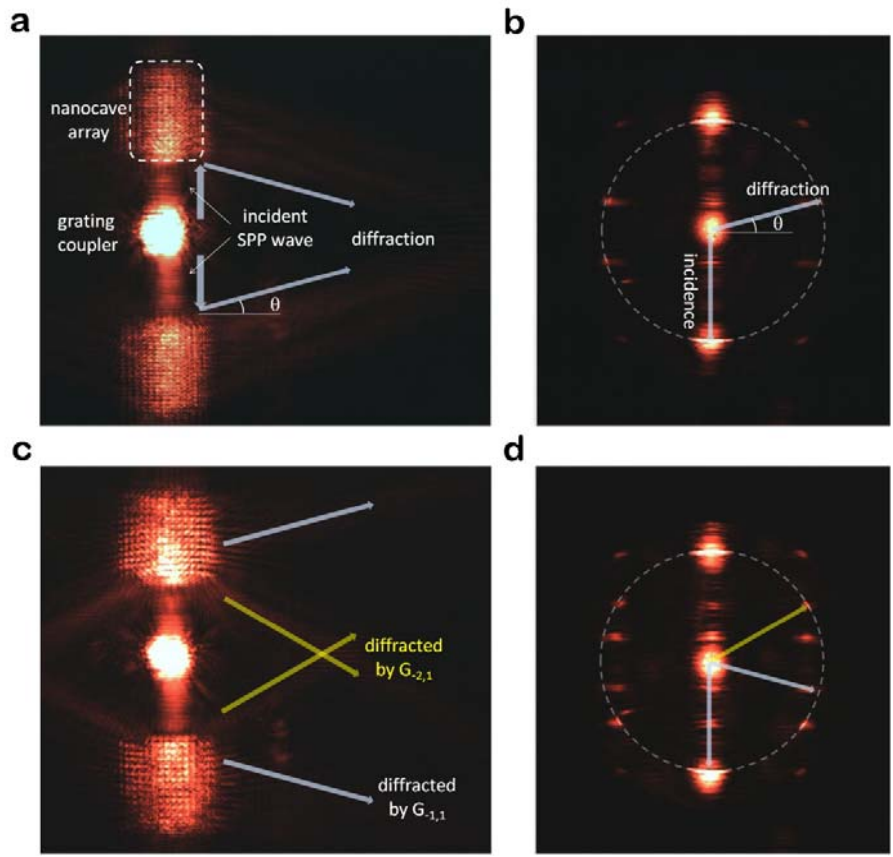

**Figure S2:**

(a), (c) Real space image of SPP propagations recorded by CCD via LRM system for sample $P_x$ =390nm (a) and $P_x$ =790nm (c), and the corresponding *k*-space patterns imaged in the Fourier plane (b) and (d), respectively. Two identical nanocave arrays are designed with mirror symmetric to the coupling grating, so as to make a precise measurement of the beaming angles.

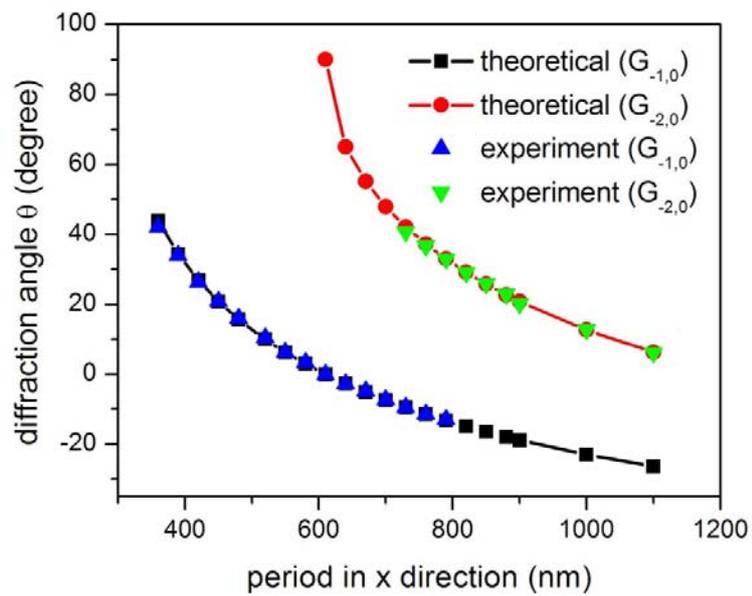

**Figure S3:**

Diffraction angles of all samples with different $P_x$ measured from the Fourier plane in LRM system, in comparison with the theoretical ones. They are indicated by different symbols.